# Compressive Sampling with Known Spectral Energy Density for SFCW-GPR Applications

Andriyan Bayu Suksmono
*School of Electrical Engineering and Informatics, Institut Teknologi Bandung*
*Jl. Ganesha No.10, Bandung, Indonesia*
`suksmono@yahoo.com, suksmono@radar.ee.itb.ac.id`

*Abstract*—A method to improve $l_1$ performance of the CS (Compressive Sampling) for A-scan SFCW-GPR (Stepped Frequency Continuous Wave-Ground Penetrating Radar) signals with known spectral energy density is proposed. Instead of random sampling, the proposed method selects the location of samples to follow the distribution of the spectral energy. Samples collected from three different measurement methods; the uniform sampling, random sampling, and energy equipartition sampling, are used to reconstruct a given monocycle signal whose spectral energy density is known. Objective performance evaluation in term of PSNR (Peak Signal to Noise Ratio) indicates empirically that the CS reconstruction of random sampling outperform the uniform sampling, while the energy equipartition sampling outperforms both of them. These results suggest that similar performance improvement can be achieved for the compressive SFCW (Stepped Frequency Continuous Wave) radar, allowing even higher acquisition speed.

## I. INTRODUCTION

Compressive sampling (CS) is an emerging method with various practical applications [1], [2]. In contrast to the Shannon sampling theorem that put a minimum limit at $2\Delta\omega$ sampling rate for a $\Delta\omega$ bandlimited signal, the CS capable to reconstruct the signal exacly based on much lower rate or fewer number of samples.

Currently, there have been efforts to improve the performance of CS by incorporating prior knowledge. Paper [3] proposes a method for sparse signal recovery that outperforms standard $l_1$ in term of fewer number of required samples. The algorithm solves a sequence of $l_1$ minimization problem where the weights used for the next iteration are computed from the value of the current solution. Related to this method, the authors of paper [4] propose an algorithm to recover sparse signal from system of underdetermined linear equations when there is prior information about the probability of each entry of the unknown signal being nonzero. While in [5], a method of modifying CS for problem with partially known support is presented.

This method is closely related to CS with partially known support described in [5]. In parctice the user is more interested to know how the modification of his/her measurement protocol improves the performance. This paper shows that a simple method to select the location of the samples in the projection domain significantly improve the objective performance for a given sample number.

The problem can be formulated as follows: given the spectral energy distribution of an A-scan GPR signal and a restricted budget on the number of measurements, how to select a set of samples that best represents the signal in the sense of CS? This problem occurs especially in the compressive SFCW (Stepped-Frequency Continuous Wave) radar [6]. It should be noted that the knowledge on absolute values of the signal's Fourier coefficients defining the spectral energy density cannot be used directly to recover the signal without any knowledge on their phase values.

In an SFCW radar, an impulse is not-directly transmitted in time-domain. Instead, the Fourier coefficients representing the signal is collected by measuring the responds of the observed objects on a range of frequency. The A-scan, which is reflections of the attenuated and shifted impulses, usually can be represented as derivative of Gaussian function. Since shifting in time domain is equivalent to shifting the phase of the Fourier coefficients in frequency domain, the magnitude of the signal spectrum will almost remain the same. Therefore, the information of the signal's spectral energy density can be used as a prior knowledge in the reconstruction. If the number of required samples can be reduced by the proposed method, the acquisition speed of the compressive SFCW radar can be increased significantly.

In this paper, one dimensional UWB (Ultra Wide Band) signal consisting of shifted and attenuated monocycles as a case, which can be generalized into higher dimensions is used. The objective performance of $l_1$ reconstruction for three different sampling schemes, namely, the random sampling, the frequency equipartion sampling (FES), and the energy equipartition sampling (EES) are compared. It has been shown in [7] that the EES performs better for direct FFT inversion representing the $l_2$ reconstruction, compared to the uniform sampling scheme. The proposed method is actually the $l_1$ extension of this scheme.

The rest of the papers is organized as follows. Section II explains briefly the principle of the standard CS and the modified CS when prior is known. In Section III, an algorithm to select a set of best samples in frequency domain for a given spectral energy density is derived. Experiments and analysis is given in Section IV and Section V concludes the paper.

## II. THEORY OF CS AND MODIFIED-CS WITH PRIOR

In the CS, reconstruction of a signal $\vec{s}$ that is sparse in a bases system $\Psi$ requires just a small number of measured samples $\vec{S}$. This subsampling process can be represented as a projection by an $M \times N$ measurement matrix $\Phi$, where

$M<<N$. Therefore, the observable $\hat{S}$, which is a subset of $\vec{S}$, can be expressed as follows

$$\hat{S} = \Phi \cdot \Psi \cdot \vec{s} = \Delta \cdot \vec{s} \qquad (1)$$

The newly defined matrix $\Delta \equiv \Phi\Psi$ represents an over-complete basis.

Equation (1) expresses an *underdetermined* system of linear equations where the number of unknown is larger than the number of the equations whose coefficients are listed in $\Delta$, therefore the solution will be non-unique. To solve this equation, CS assumes that the signal is sparse, which means that the number of the $\Psi$-domain coefficients, i.e.

$$\|\vec{s}\|_0 \equiv \sum_{n=1}^{N} |s_n|^0 \qquad (2)$$

is the smallest one. Actually, minimization of (2) is a combinatorial problem that computationally intractable. When the signal is highly sparse, the solution of (2) for $L_0$ is identical to the solution of a more tractable $L_1$ problem [7], [8], by minimizing

$$\|\vec{s}\|_1 \equiv \sum_{n=1}^{N} |s_n|^1 \qquad (3)$$

In fact, minimization of (3) can be recast as a convex programming problem [11], [12], whose solvers are widely available, such as the Interior Point Method.

An important issue regarding this solution is that $\Phi$ and $\Psi$ should be sufficiently incoherent. The measure of coherence between two bases $\mu(\Phi,\Psi)$ is defined as [13]:

$$\mu(\Phi,\Psi) = \max_{\phi \in \Phi, \psi \in \Psi} |\langle \phi, \psi \rangle| \qquad (4)$$

where $\phi$ and $\psi$ are column (row) vectors of $\Phi$ and $\Psi$, respectively.

It has been shown that a general random basis has a high degree of incoherence with any basis, including the identity or spike bases **I**. Therefore, one can choose a random matrix as the projection bases $\Phi$. In such a bases, the number of required sample $K$ is [13]

$$K \geq C \cdot \mu^2(\Phi,\Psi) \cdot F \cdot \log(N) \qquad (5)$$

where $C$ is a small constant, $F$ denotes the degree-of-freedom of the signal or the number of non-zero coefficient of the signal when represented in the sparsity bases $\Psi$.

For a suitable number of measured data $K$ given by (5), CS guarantees to recover perfectly the time domain signal through optimization

$$\min_{\vec{s} \in R^N} \|\vec{s}\|_1 \; s.t. \; \hat{S}_k = \langle \bar{\varphi}_k, \Psi\vec{s} \rangle, \forall k \in \{1,2,...,K\} \qquad (6)$$

where $\bar{\varphi}_k$ is a row vector of $\Phi$. In brief, the CS principle states that for a small, but sufficient, number of observations, it is possible to recover a sparse signal $\vec{s}$ from its subsamples $\hat{S}$ through $L_1$ optimization given by (6).

The performance of CS can be improved when there is (are) prior information of the signal. In [4] and [5], weights are elaborated into the formulation of the optimization, i.e., equation (6) is modified into

$$\min_{\vec{s} \in R^N} \|W\vec{s}\|_1 \; s.t. \; \hat{S}_k = \langle \bar{\varphi}_k, \Psi\vec{s} \rangle, \forall k \in \{1,2,...,K\} \qquad (7)$$

where **W** is a diagonal matrix of positive weights. On the other hand, when the support $T$ of the signal is known, one can also improve the performance as suggested in [5] by reformulating (6) into

$$\min_{\vec{s} \in R^N} \|(\vec{s})_{T^C}\|_1 \; s.t. \; \hat{S}_k = \langle \bar{\varphi}_k, \Psi\vec{s} \rangle, \forall k \in \{1,2,...,K\} \qquad (8)$$

where $T^C$ denotes the complement of $T$.

Compared to the (re-) weighted and the known-support modified CS, the proposed method uses the prior in a slightly different manner. When the absolute of the spectrum or the distribution of signal energy in frequency domain is known, the method selects only particular samples that follow the distribution of the spectral energy density. The detail scheme is described in the following Section.

## III. FREQUENCY DOMAIN SAMPLING AND THE PRINCIPLES OF EQUIPARTITION

In uniform sampling, the frequency band is divided into $N$ sub-bands uniformly, i.e,

$$\Delta\Omega_1 = ... = \Delta\Omega_i = ... = \Delta\Omega_N \qquad (9)$$

Such a trivial scheme will be named as the frequency equipartition sampling (FES). In this method, a different approach to get a better time-domain reconstruction results is proposed; i.e., by proportionally counting the contribution of the spectral energy in each frequency sub-bands, which is illustrated in Fig.1. The left part of Fig.1 shows a time-domain impulse $s(t)$, a monocycle for example, while the right part is its spectral energy density $|S(\Omega)|$ obtained from the Fourier transform of $s(t)$.

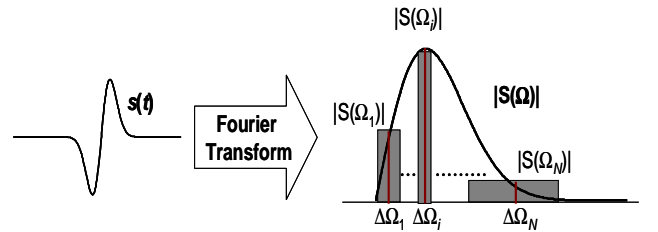

Fig.1 Non uniform frequency spacing scheme based on equipartition of the spectral energy

The main idea in the new scheme is to select sub-bands of frequencies and its range $[\Omega_i - \Delta\Omega_i/2, \Omega_i + \Delta\Omega_i/2]$, so that the energies in the $\Delta\Omega_i$ intervals are identical. It is shown in the figure as dashed bars that have identical areas. The centre of the sub bands {$\Omega_i$} will become the location of selected samples in frequency domain.

```
The EES Algorithm

1. For a given spectral energy
   distribution |S(Ω)|, define the
   frequency range [Ω_L, Ω_U] and the
   number of sample N.
2. Calculate the total sum of spectral
   energy Ê = ∫_{Ω_L}^{Ω_U} |S(Ω)|dΩ and the average
   energy in the subband ε ≡ Ê/N .
3. Starting from the lowest to the
   highest frequency:
   a. Integrate E(Ω) over an interval
      ΔΩ such that the total energy in
      the interval equal to ε. The
      middle of the interval is the
      location of selected sample.
   b. Repeat Step 3.a until all of the
      sampling points in the set
      {Ω_i|i=1, …, N} are found.
```

Fig.2 The EES Algorithm to select sampling points

For a given working frequency band bounded by $\Omega_L$, and $\Omega_U$, approximation of the absolute magnitude sum or the magnitude-energy in each dashed area, $\varepsilon$, is given by:

$$\varepsilon \equiv \frac{\hat{E}}{N}, \text{ where } \hat{E} = \int_{\Omega_L}^{\Omega_U} |S(\Omega)|d\Omega \quad (10)$$

A new scheme called the equipartition of the energy sampling (EES) divides the full band into $N$ subbands whose magnitude energy are equals.

$$\Delta\Omega_1|S(\Omega_1)| = ... = \Delta\Omega_i|S(\Omega_i)|$$
$$= ... = \Delta\Omega_N|S(\Omega_N)| \quad (11)$$

Consequently, the $i$-th frequency $\Omega_i$ is obtained, the corresponding range of frequency is

$$\left(\Omega_L + \sum_{k=1}^{i}\Delta\Omega_k - \frac{\Delta\Omega_i}{2}\right) \leq \Omega_i < \left(\Omega_L + \sum_{k=1}^{i}\Delta\Omega_k + \frac{\Delta\Omega_i}{2}\right) \quad (12)$$

and the width of the $i$-th subband is

$$\Delta\Omega_i = \varepsilon/|S(\Omega_i)| \quad (13)$$

According to (12) and (13), to determine the set of frequencies {$\Omega_i$} one needs the spectral energy density |S(Ω)| and the number of sample $N$. A simple algorithm to determine the sample locations in frequency domain according to energy equipartition sampling (EES) can be immediately formulated. Figure 2 displays the EES algorithm.

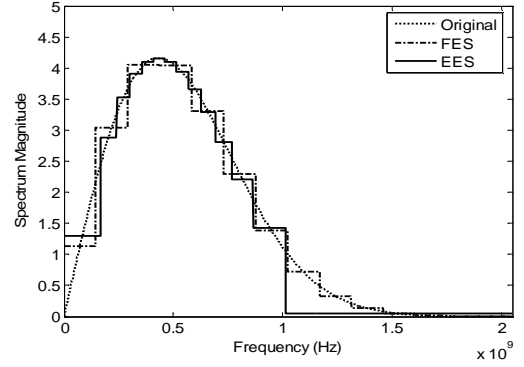

Fig.3. Comparison of EES with FES for a 14 samples

An illustration of EES compared to the FES for a 2GHz impulse of monocycle signal divided into 14 sub-bands is presented in Fig.3. The selected sample is located in the centre of each subband for corresponding method. The figure shows that the subband becomes wider when the spectral energy density is lower, yields non-uniformly distributed frequency-domain sample positions.

## IV. Experiments and Analysis

In the experiment, a 256 length discrete time signal representing a 2 GHz monocycle signal of the GPR A-scan is generated. The A-scan will consisting of shifted and attenuated monocycle impulses, depending on the number of reflections and their range or distance from the antenna. In CS terminology, the number of the impulse defines the DoF (Degree of Freedom) or the sparsity of the signal. Therefore, minimum number of required samples given in (5) will change according to the value of DoF. For the present case, one and three random reflections are simulated.

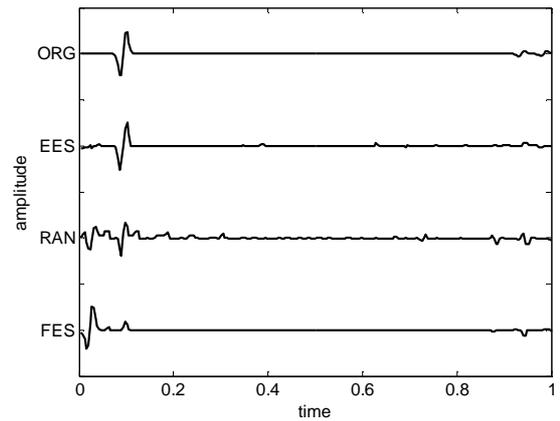

Fig.4 Simulation results for a 14 samples for three sampling schemes: EES, random sampling, and FES

Figure 4 shows the reconstruction results of monocycle signal based on 14 samples selected by three sampling methods. The top part shows the original signal, while the next ones are reconstructed signal based on samples obtained by the EES, random sampling, and the FES, subsequently. The PSNR values of reconstructed signal for the present case by the FES method is -12.9 dB, random sampling gives -6.2 dB, and the EES yields 19.6 dB. Therefore, the EES gives the best results compared to both of the random sampling and the FES.

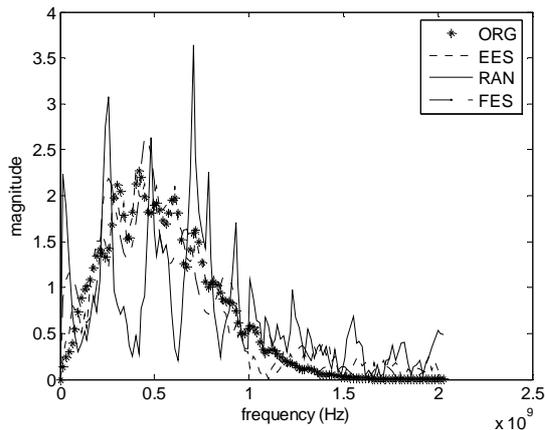

Fig.5 The spectrum of the original and reconstructed signals

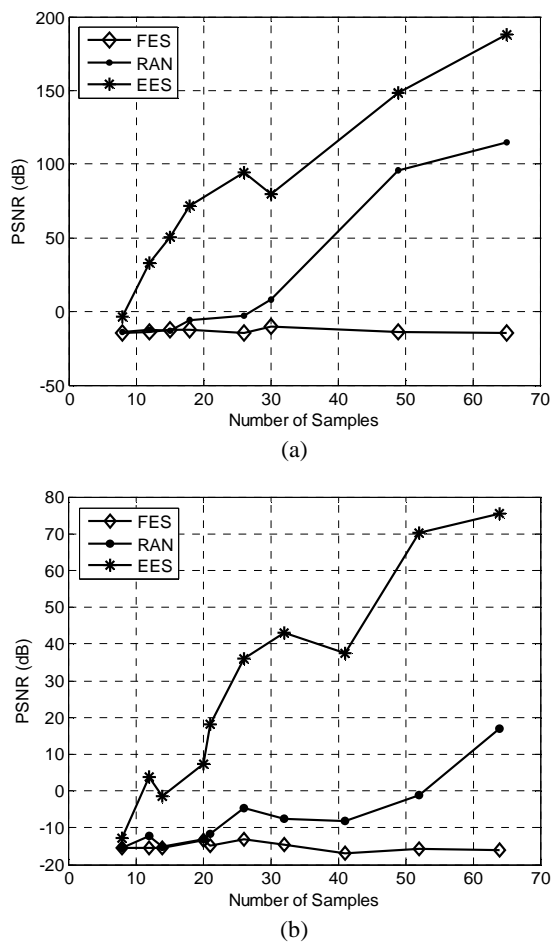

Fig.6 The $l_1$ reconstruction performance in term of PSNR for various number of samples with three sampling methods: frequency equipartition sampling (FES), random sampling (RAN), and energy equpartition sampling (EES). Figure (a) shows the performance of a signal with one monocycle, and (b) with three shifted and attenuated monocycles

Figure 5 shows the spectrum of original signal and the reconstructed ones. The spectrum also shows that the reconstructed signal from EES best fits the original magnitude spectrum.

Figure 6 shows PSNR performance of various numbers of samples with (a) one and (b) three DoF (attenuated and shifted monocycles). Each data point is an average of seven times signal generation, sampling, and reconstruction. This figure shows that the EES consistently outperforms both of the FES and random sampling in term of PSNR and demonstrates that higher DoF requires more sample to achieve the same PSNR as the lower one.

V. CONCLUSIONS

A new method to improve $l_1$ reconstruction in CS with known spectral energy density is described. Performance of three sampling schemes, i.e, the FES, random sampling, and EES are compared an analized. It is shown that the EES outperforms both of the random sampling and FES. This result enables a possibility of CS imaging with much fewer number of samples, hence higher acquisition speed, than suggested by random sampling in the standard CS method.

ACKNOWLEDGEMENT

This work is supported by the ITB Grant of Research Division (Riset KK-ITB) 2009.